\documentclass[preprint,5p,times,twocolumn]{elsarticle}

\usepackage{amssymb}
\biboptions{sort&compress}

\usepackage{amsmath,amsfonts,amssymb,amsthm}
\usepackage{amssymb}
\usepackage{mathtools}
\usepackage{commath}
\usepackage{comment}
\usepackage{caption}
\usepackage{subcaption}
\usepackage[sc,osf]{mathpazo}
\usepackage[english]{babel}
\usepackage{breqn}
\usepackage{array}

\usepackage[dvipsnames]{xcolor}
\usepackage{graphicx}
\newtheorem{theorem}{Theorem}

\newtheorem{definition}{Definition}
\newproof{pot}{Proof}
\newtheorem{remark}{Remark}
\newtheorem{lemma}{Lemma}
\newtheorem{corollary}{Corollary}

\newcommand{\PP}{\mathcal{P}}
\newcommand{\LL}{\mathcal{L}}
\newcolumntype{C}[1]{>{\centering\let\newline\\\arraybackslash\hspace{0pt}}m{#1}}
\usepackage{float}

\journal{Systems \& Control Letters}

\begin{document}

\begin{frontmatter}

\title{Safe Region Multi-Agent Formation Control With Velocity Tracking}

\author[l1]{Ayush Rai}
\ead{rai29@purdue.edu}

\author[l1]{Shaoshuai Mou\corref{mycorrespondingauthor}} \cortext[mycorrespondingauthor]{Corresponding author}
\ead{mous@purdue.edu}

\affiliation[l1]{organization={School of Aeronautics and Astronautics},
            addressline={Purdue University}, 
            city={West Lafayette},
            state={IN},
            country={USA}}

\begin{abstract}
This paper provides a solution to the problem of safe region formation control with reference velocity tracking for a second-order multi-agent system without velocity measurements. Safe region formation control is a control problem where the agents are expected to attain the desired formation while reaching the target region and simultaneously ensuring collision and obstacle avoidance. To tackle this control problem, we break it down into two distinct objectives: safety and region formation control, to provide a completely distributed algorithm. Region formation control is modeled as a high-level abstract objective, whereas safety and actuator saturation are modeled as a low-level objective designed independently, without any knowledge of the former, and being minimally invasive. Our approach incorporates connectivity preservation, actuator saturation, safety considerations, and  lack of velocity measurement from other agents with second-order system dynamics which are important constraints in practical applications. Both internal safety for collision avoidance among agents and external safety for avoiding unsafe regions are ensured using exponential control barrier functions. We provide theoretical results for asymptotic convergence and numerical simulation to show the approach's effectiveness.

\end{abstract}


\end{frontmatter}


\section{Introduction}
\label{sec:introduction}
Formation control (FC) in multi-agent systems has attracted a significant amount of research attention 
\cite{Anderson2008, Ren2010,  Oh2015, chen2015controllability, chen2015time, de2016distributed, yang2018distributed}, 
which aims to coordinate a group of agents to form a certain geometric shape only using local information/measurements between agents and their nearby neighbors. It has been widely used in various fields to achieve collective behavior, distributed sensing, cooperative manipulation, and other applications. Classical FC has been widely studied including but not limited to stabilization control of rigid directed and undirected formations \cite{sun2016exponential,sun2017distributed}, maneuvering and formation tracking \cite{de2016distributed,yang2018distributed}, and controllability of formations \cite{chen2015controllability,chen2015time}. Based on the information used to define and maintain the multi-agent formations, the techniques developed for FC can be broadly categorized as follows: position-based FC \cite{Ren2007, Dong2008}, displacement-based FC \cite{Olfati-Saber2004, Ren2005}, distance-based FC \cite{sun2016exponential,sun2017distributed,SMA16TAC,SMA15Auto}, and angle-based FC \cite{Basiri2010,Ming22TAC_angle,Ming23_TACAngle,Ming23_TAEAngle}. These approaches define desired formations based on global positions, displacement (i.e., relative position vectors between agents and their neighboring agents in agents' local coordinate systems), distances between agents and their nearby neighbors, and angle information, respectively.

When implementing these distributed algorithms for multi-agent FC in practice, several practical constraints naturally arise, besides controlling a multi-agent formation in its desired geometric shape. First, a multi-agent formation is usually required to be driven to reach a specific region from its mission requirement, as pointed out in \cite{cheah2007region,cheah2009region,miao2019multi,yu2019neural, yang2021finite} giving rise to the problem of region formation control (RFC).
The region-based shape controller developed in \cite{cheah2009region} can drive all agents to a specific region but only maintain a minimum distance among agents rather than a desired formation shape. A neural network-based approach is presented in \cite{yu2019neural}, which successfully achieves the objective of formation control within a specified region. However, this method entails the training of the network and places demand on actuators that may be relatively high during the initial phases of the region-tracking process. Second, agents need to avoid collisions with other agents and obstacles while converging and maintaining the formation, i.e. safe formation control (SFC). The work \cite{deghat2011safe} employs a distributed consensus algorithm for obstacle avoidance in the assumption that a leader can never encounter an obstacle and the two agents in the neighborhood cannot have active collision course flags simultaneously. Hierarchical control approaches have been developed for SFC in \cite{regula2012formation,regula2014formation} by introducing a very nice path generator to generate safe paths at the cost of repeatedly constructing graphs to evaluate the risk of collisions. Several methods leverage artificial potential functions \cite{peng2019output, yu2019neural} to integrate safety considerations. However, these approaches may exhibit conservative performance in RFC since potential functions are not minimally invasive. Very recently the authors of  \cite{zhang2022barrier} have developed a distributed MPC based on reinforcement learning integrated with a barrier function, which successfully solved the problem of SFC at the cost of high computational complexity for the training process. Third, as highlighted in surveys \cite{Oh2015, zavlanos2011graph}, maintaining connectivity in the multi-agent systems is a common assumption in many existing algorithms for multi-agent FC, which is however not valid in practice because of agents' limited sensing capability. Connectivity preservation in displacement-based FC was studied in \cite{Ji2007} where the desired formation is achieved for single-integrator modeled agents. Another way through which connectivity has been preserved in the literature is by maximizing the second smallest eigenvalue of the Laplacian of graph network \cite{Zavlanos2011}. Although there has been significant progress achieved for multi-agent formation control, there lacks a distributed algorithm for multi-agent formation control with consideration of all the above practical constraints, i.e. region reaching, collision avoidance, and connectivity preservation.

In this paper, we aim to develop a unified distributed algorithm for multi-agent safe region formation control (SRFC), i.e. steering the multi-agent system to converge to a desired formation in a specific region with collision avoidance, velocity tracking, actuator saturation, and connectivity preservation. We consider second-order dynamics without velocity measurements of other agents which significantly reduces the communication requirements. In contrast to existing literature, the key contributions of the proposed work are as follows. First, we provide a unified distributed algorithm that accounts for all the practical considerations and has not directly been solved using methods developed in \cite{cai2014multi, zhang2015three, deghat2015combined,sun2017rigid, hou2009dynamic,chen2017connection, yang2018connectivity, peng2019output,yu2019neural}. Second, unlike most of the work \cite{peng2019output,yu2019neural}, we model safety as an inherent feature of the system by using control barrier functions to provide more flexibility to the system. The absence of velocity or control input measurements makes it a challenging task to design such a barrier function in completely a distributed fashion. Third, our approach can be easily extended to both region tracking and perimeter monitoring problems. Finally, the work distinctively shows the significance and developments of each component of the control input for the respective objective. Our approach is well supported by theoretical results and numerical simulations showcasing the effectiveness of the proposed approach.

The overall strategy adopts a modular design based on \cite{wieland2007constructive}, where the feedback law is designed to ensure the safety of the multi-agent system (\textit{low-level objective}) at all  times while providing maximum flexibility for the abstract objectives (\textit{high-level objective}), making safety as an intrinsic part of the system. We break down the objective of achieving safe region formation control (SRFC) into two distinct components. At the high level, we focus on maintaining the desired formation shape maneuvering towards a target region, and velocity tracking. At the low level, our objective is to address collision avoidance and actuator saturation. To achieve these objectives, we integrate a nominal control approach based on an artificial potential function for formation shape, consensus algorithms for connectivity preservation, and the concept of control barrier functions (CBF) as proposed in \cite{Ames2019} for collision avoidance. We also use the concept of \textit{stealthy} leader/leaders (partial access to group reference state) which are agents assumed to have access to information about the target region and reference tracking velocity. They are stealthy in the sense that they have the same dynamics and cannot be distinguished from other agents, providing a certain degree of security. The proposed algorithm ensures connectivity preservation in a distributed way without any knowledge of the total number of agents or any estimation strategy for the second smallest eigenvalue of the Laplacian matrix of the network as used in works like \cite{poonawala2014collision}. 

The remainder of the article is organized as follows. We present our problem formulation in Section 2. The main results are outlined in Section 3, which comprises two sub-sections. Section 3.1 delves into the nominal control approach, while Section 3.2 offers insights into the implementation of safe control and includes a concise introduction to control barrier functions (CBFs). In Section 4, we present simulation results that demonstrate the efficacy of the developed control strategy for a range of scenarios and choices of hyper-parameters. Finally, Section 5 presents the conclusion and future scope of this work.

\textit{Notations:} We use the notation $\mathbb{R}^n$ to denote the set of all $n \times 1$ real vectors. The transpose of a matrix or vector is denoted by $(\cdot)^{T}$. The $L_2$ norm of a vector is denoted by $\|.\|$, whereas the infinity norm of a vector is denoted by $\|.\|_\infty$. For any set $(\cdot)$, $\partial(\cdot)$ denotes the set of all the points lying on the boundary of the set $(\cdot)$.

\section{Problem Formulation} \label{Sec_Prob}
Consider a multi-agent system composed of $N$ agents, with each agent $i$'s dynamics modeled by the following double-integrator:
\begin{align}
    \Dot{p}_i(t) &= v_i(t), \nonumber\\
    \Dot{v}_i(t) &= u_i(t),
    \label{eq:dynamics}
\end{align}
in which $p_i \in \mathbb{R}^2$, $v_i \in \mathbb{R}^2$, and $u_i \in \mathbb{R}^2$ denote the position, velocity, and control input of agent $i$, respectively. All agents are subject to input saturation constraint as $\|u_i\|_{\infty} \leq u_{max}$. 

Each agent has a limited sensing region defined by a closed hyper-sphere with a positive radius $r$ centered at the agent’s current position. Hence, no information about the space beyond this closed disk can be obtained by the agent. Given the sensing radius, $r>0$, for any $\epsilon \in (0,r)$ and $t \geq 0$, the neighbor set of agent $i$, $\mathcal{N}_i(t)$, at any time instant is defined in a hysteresis fashion as follows
\begin{itemize}
    \item $\mathcal{N}_i(0) = \left\{j | \|p_i(0) - p_j(0)\| < r-\epsilon, j \in \{N\} \right\}$
    \item If $\|p_i(t) - p_j(t)\| \geq r$, then $j \notin \mathcal{N}_i(t)$
    \item If $j \notin \mathcal{N}_i(t^-)$ and $ \|p_i(t) - p_j(t)\|  < r-\epsilon$, then $j \in \mathcal{N}_i(t)$ 
\end{itemize}
where $\|p_i(t) - p_j(t)\|$ denotes the distance between agent $i$ and its neighbor $j$ at time $t$, whereas $t^-$ denotes any time instance before $t$. This allows us to define a communication network formed by the multi-agent system using an undirected dynamic graph $\mathcal{G}(t) = (\mathcal{V},\mathcal{E}(t))$, where the node set $\mathcal{V} = {1,2,...,N}$ denotes the set of agents and the edge set $\mathcal{E}(t) \subseteq \mathcal{V}\times\mathcal{V}$ denotes the communication link between an agent and its neighbors at any time $t$.

\begin{remark} The above hysteresis-based approach to describe agents' neighbors is adapted from \cite{Su2010}. Under these conditions, in order for two agents, which are not  neighbors at time $t^-$,  to be neighbors at any time $t$, the distance between them will be less than $r-\epsilon$. Further, if they are already neighbors then for them to lose connectivity, the distance between them should be greater than $r$.
\end{remark}

\begin{definition}
Desired formation: Consider a network of $N$ agents given by $\PP^* = \{p_1^*,\dots, p_N^*\}, p_i^* \in \mathbb{R}^2$ representing a feasible target shape, such that graph formed by $\PP^*$; $\mathcal{G}^* = (\mathcal{V},\mathcal{E}^*)$, where $$ \mathcal{E}^* = \left\{(i,j) | \|p_i^* - p_j^*\| < r-\epsilon, i,j \in \mathcal{V} \right\}.$$ The desired formation is defined as:
\[\PP_d = \{p_i | (p_i-p_j)=(p_i^*-p_j^*) \; \forall \; j \in \mathcal{N}_i; \; i \in \mathcal{V}\}.\]
Therefore, $\PP_d$ represents the collection of all formations that are identical to $\PP^*$ in terms of shape and size but may differ in translation.  
\end{definition}

Besides maintaining a desired geometric shape, the multi-agent system will be steered to a \emph{ target region}, assumed convex, denoted by $\Omega \subset \mathbb{R}^2$ while avoiding $K$ \emph{unsafe regions} denoted using $\overline{\Omega} = \{\overline{\Omega}_k \; | \; \overline{\Omega}_k \subset \mathbb{R}^2 \; \forall\; k = 1,2,\dots, K\}$. A collision involving agent $i$ and obstacle $k$ is defined as occurring when the position of agent $i$ lies within the boundary of the obstacle $\overline{\Omega}_k$, i.e. $p_i \ \in \overline{\Omega}_k$. On the other hand, an inter-agent collision between two agents is defined as occurring when the distance between them is less than a specified threshold value, denoted as $\delta_{in}$. Similar to \cite{ren2008consensus, Abdessameud2010OnConstraints}, we let $v_d(t)$ denote the reference velocity for all agents to converge to after entering the target region, although only some agents know $v_d(t)$, which are referred as stealthy leaders.  

\begin{definition}
Stealthy leader/leaders: The set of stealthy leaders $\mathcal{L} \subseteq \mathcal{V}$ is defined as the agents having the information about the final target region and the reference tracking velocity signal.
\end{definition} 

We make the assumption that only stealthy leaders possess the ability to identify the final target region and possess information about the constant reference tracking velocity signal. However, in the case of a time-varying tracking velocity, all agents must be aware of its derivative for the formation to asymptotically track the reference tracking velocity. Additionally, we assume that the agents have the capability to sense the relative positions of other agents or detect obstacles within their sensing range. In our setup, it is not necessary for all agents to be familiar with the global coordinate system, except for the stealthy agents. Each agent is presumed to be aware of its local coordinate system, with orientations aligned with that of the global coordinate system.

The  objective  of  this work is  to  devise  a  distributed control strategy for a multi-agent system, to reach a final target region, attain a desired formation, avoid  all  the  unsafe  regions  in  the  space, and track the dynamic velocity reference signal after reaching the final target region. The objectives can be expressed mathematically as follows: Given the system dynamics \eqref{eq:dynamics} and the constraint of input saturation, the aim is to design a distributed control input that
\begin{enumerate}
    \item Attains desired formation:
    \begin{equation}
    \lim_{t\to\infty} \left( p_i(t) - p_j(t) \right) = p_i^* - p_j^* \; \forall \; i,j \in \mathcal{V}.
    \label{eq:obj_formation}
    \end{equation}
    
    \item Reaches target region:
    \begin{equation}
    p_i(t_f) \in \Omega  \; \forall \; i \in \mathcal{L},
    \label{eq:obj_region}
    \end{equation}
    where $t_f$ is defined as the time required for the stealthy leader to reach the target region $\Omega$ from its initial position.

    \item Tracks reference velocity signal $v_d$ after reaching the target region:
    \begin{equation}
    \lim_{t\to\infty} v_i(t)=v_j(t)=v_d(t) \;; \;t>t_f.
    \label{eq:obj_tracking}
    \end{equation}
    
    \item Ensures internal and external safety at all times, for given $\delta_{in}>0$:
    \begin{subequations}
    \begin{align}    
    p_i(t) &\notin  \overline{\Omega}_k \; \forall \; i \in \mathcal{V}, \; k = 1,\dots,K, \\
    ||p_i(t) &- p_j(t)|| \geq \delta_{in} \; \forall \; i,j \in \mathcal{V} ; \; i\neq j.
    \end{align}
    \label{eq:obj_safety}
    \end{subequations}
\end{enumerate}

\begin{remark}
We note that the target-reaching objective is defined only for the stealthy leader and not for all the agents. However, to ensure this objective for all agents, one could introduce a new convex target region $\Omega_{new}$ that is at least $\Delta$ smaller compared to the original region $\Omega$, i.e. $ \Omega_{new} = \{y| \|x,y\|\geq \Delta \forall x\in \partial\Omega, y \in \Omega \} $, where $\Delta$ represents the maximum distance between any two agents in the graph $\mathcal{G}^*$.
\end{remark}

\section{Main Algorithms and Results} \label{sec:main_algorithm}

In this section, we propose a distributed algorithm for multi-agent formation control that aims to achieve the objectives stated in equations \eqref{eq:obj_formation}-\eqref{eq:obj_safety}. To accomplish this, we decompose the overall objective into two sets of tasks. The first set (Section 3.1), represented by equations \eqref{eq:obj_formation}-\eqref{eq:obj_tracking}, focuses on high-level objectives and involves the development of a nominal control approach. The second set (Section 3.2), represented by equation \eqref{eq:obj_safety}, pertains to a lower-level task  designed as an inherent capability of the system and incorporates the concept of control barrier functions with the nominal control. Motivated by \cite{ren2008consensus,Abdessameud2010OnConstraints}, the proposed control law for the SRFC for agent $i$ is given as a solution to QP problem:
\[ \quad
u_i^*=\arg \min_{z_i} \frac{1}{2} \norm{z_i- \bar u_i}^2,
\]
such that
\begin{align*}
    A_{ik}+B_{ik}^Tz_i & \geq 0 \; \forall \; i \in \mathcal{V}, \; k = 1,2,\dots, K, \\
    \frac{1}{2} \bar A_{ij}+ \bar B_{ij}^Tz_i &\geq 0 \; \forall \; j \in \mathcal{N}_i; \; j\neq i; \;  i \in \mathcal{V}, \\
    \|z_i\|_{\infty}  &\leq  u_{max} \; \forall \; i \in \mathcal{V} ,
\end{align*}
where $\bar u_i$ is the nominal control of the agent solving the high-level objectives. The definitions $A, B, \bar A,$ and $\bar B$ are discussed later in Section \ref{sec:safe_control}. The nominal control is given by
\begin{equation}
\begin{split}
& \bar u_i = \mathbb{1}_{\Omega} \dot{v}_d-c_1\sum_{j \in \mathcal{N}_i}\nabla_{p_i}\Phi_{ij} -c_2\sum_{j \in \mathcal{N}_i}\hat v_{ij} \\- c_3 \mathbb{1}_{i \in \LL}&\mathbb{1}_{\Omega^-} \frac{p_i-\mathcal{P}_\Omega(p_i)}{\|p_i-\mathcal{P}_\Omega(p_i)\|} -c_4\mathbb{1}_{i \in \LL} \mathbb{1}_{\Omega}\tanh(c_5(p_i-\gamma_i)),
\label{eq:nom_control}
\end{split}
\end{equation}
where $\mathbb{1}_{i \in \LL}$ is an indicator function whose value is $1$ if agent $i$ is a stealthy leader, $\mathbb{1}_{\Omega}$ and $\mathbb{1}_{\Omega^-}$ are indicator functions corresponding on whether the agent is inside the target region or not. Here, $\Phi_{ij}$ represents the artificial potential function, which is defined later in Section 3.1, and $\mathcal{P}_\Omega(p_i)$ refers to the projection of the point $p_i$ onto the set $\Omega$. The estimate of the relative velocity vector $\hat v_{ij}$ is defined using an auxiliary variable $\phi_{ij}$ as
\begin{align}
    \hat v_{ij} = -\eta ( \phi_{ij} - (p_i-p_j)), \label{eq:vel_estimate}\\
    \dot \phi_{ij} = -\eta ( \phi_{ij} - (p_i-p_j)), \label{eq:aux1_def}
\end{align}
and auxiliary variable $\gamma_i$ is defined as
\begin{equation}
    \dot{\gamma}_i= v_d + c_5(p_i-\gamma_i).
    \label{eq:aux2_def}
\end{equation}

Here  $c_1$, $c_2$, $c_3$, $c_4$, $c_5$, and $\eta$  are strictly positive, $\phi_{ij}(0)$ is initialized with $(p_i(0)-p_j(0))$, whereas $\gamma_i(0)$ can take any arbitrary value. 

\subsection{Region Formation Control without Safety} 
\label{sec:main_without_safety}
We first develop a nominal control without consideration of safety. By the integration of artificial potential function, distributed consensus, and projection to convex sets, we propose the control law for agent $i$ with dynamics (\ref{eq:dynamics}) in the multi-agent system in the following form:
\begin{equation}
\bar u_i = \mathbb{1}_{\Omega} \dot{v}_d + c_1 u_{i1} + c_2 u_{i2} + c_3 u_{i3} + c_4 u_{i4}.
\end{equation}
Here $u_{i1}$ represents the gradient of the artificial potential function employed to maintain connectivity and guide the system toward the desired formation. Velocity consensus is ensured with $u_{i2}$ by constructing an estimate of the relative velocity, guaranteeing that the formation moves together. The term $u_{i3}$ involves projection onto the convex target set, aiming to steer the formation towards the target region $\Omega$. Lastly, $u_{i4}$ corresponds to velocity tracking.

\subsubsection{Nominal: Formation control}

Artificial potential functions have been used by many researchers for formation control and rendezvous problems in multi-agent systems. The basic idea remains the same as observed in natural phenomena like electrostatics and gravitation. We define a scalar, non-negative potential function $\Phi_{ij}$ as the measure of formation control error for each neighbor pair $(i,j)$. With $d_{ij}$ defined as $d_{ij}=p_i^*-p_j^*$, we require the potential function $\Phi$ to obey the following:
\begin{enumerate}
    \item $\Phi_{ij}$ is continuously differentiable.
    \item $ \left(\frac{ \partial \Phi_{ij}}{\partial (p_i - p_j-d_{ij})}\right)^T\left(p_i - p_j-d_{ij}\right) \geq 0$ for $\| p_i - p_j\| \in [0,r)$. The potential should increase with distance as it moves far from the desired displacement and is zero only at the desired displacement.
    \item $ \lim_{\| p_i - p_j\| \to r} \Phi_{ij} \to \infty$. Potential should be sufficiently large when the distance between the two agents reaches the sensing radius to preserve the connectivity.
\end{enumerate}
We propose the potential function $\Phi_{ij} = \frac{\| p_i - p_j -d_{ij}\|^2}{r^2-\| p_i - p_j\|^2+ \mu}$  between agents $i$ and $j$ for displacement-based formation control, where $\mu>0$ is some small positive number.   Performing gradient descent along $\Phi$ ensures that the agents remain close enough in contemplation of connectivity preservation. The minimum of this potential is achieved when $p_i-p_j=d_{ij}$, i.e. when the agents achieve the desired displacement. Based on the potential function defined, an explicit distributed control input for the system can be given as
\begin{equation}
u_{i1} = -\sum_{j \in \mathcal{N}_i} \nabla_{p_i} \Phi_{ij},
\end{equation}
where the summation is taken over all agents in the neighborhood of agent $i$. 

\begin{remark}
    Displacement-based formation control can generate much richer formations in a MAS system compared to distance or angle-based FC \cite{chen2017connection}. The desired displacements provided to the system can be changed with time to perform any rotation or scaling if required. For these reasons, we only present the analysis for displacement-based formation control in this work. The complete analysis can be easily extended for distance-based formation control using the potential function $\Phi_{ij} = \frac{(\| p_i - p_j\|^2 -\|d_{ij}\|^2)^2}{r^2-\| p_i - p_j\|^2+ \mu}$. 
\end{remark}

Due to the second-order dynamics, a distributed consensus protocol on velocity is required to ensure that the agents move together and have a stable convergence. Asymptotically we want all the agents to have the same velocity. Due to the lack of velocity measurement of the neighboring agents, we use $\hat v_{ij}$ as an estimate of the relative velocity between agent $i$ and $j$. Hence,
\begin{align*}
    u_{i2} = -&\sum_{j \in \mathcal{N}_i}\hat v_{ij},
\end{align*}
where $\hat v_{ij}$ is defined using \eqref{eq:vel_estimate} and \eqref{eq:aux1_def}.
The absence of this term leads to significant oscillatory motion, which is extensively discussed in Section 4.1.

\subsubsection{Nominal: Region formation control}

For the discussing region constraint, we define the concept of projection of a point on a convex set.

\begin{definition}
Projection on a convex set: Given a closed convex set $\mathbb{C}$, the projection of a point $z$ onto $\mathbb{C}$ is defined by the point $z^{*} = \mathcal{P}_{\mathbb{C}}(z)$ that minimizes $\|z-z^{*}\|$.
\end{definition}

We assume that at any time instant, all the stealthy leaders have the knowledge of the projection of their position on the target region. The proposed potential function is given by
\begin{equation}
u_{i3} = -\mathbb{1}_{i \in \LL}\mathbb{1}_{\Omega^-} \frac{p_i-\mathcal{P}_\Omega(p_i)}{\|p_i-\mathcal{P}_\Omega(p_i)\|},  
\label{projection_term}
\end{equation}
where $\mathcal{P}_\Omega(p_i)$ denotes the projection of point $p_i$ on the set $\Omega$. This projection function is zero when $p_i$ is in the convex set $\Omega$, and non-zero otherwise.


\subsubsection{Nominal: Region formation control with velocity tracking}

For velocity tracking we want the formation to track $v_d$ once the formation enters the target region, i.e. after time $t_f$. This velocity signal and its derivative will be governed based on the objective at hand, like surveillance, rescue, or attack. Motivated by \cite{Abdessameud2010OnConstraints}, we formulate the control law such that if the velocity signal is constant it only needs to be available to the stealthy leaders. On the other hand, if the velocity signal is time-varying then its derivative should be made available to all the agents. This gives us
\begin{equation}
u_{i4} = -\mathbb{1}_{i \in \LL} \mathbb{1}_{\Omega}\tanh(c_5(p_i-\gamma_i)),  
\label{eq:vel_tracking}
\end{equation}
where auxiliary variable $\gamma_i$ is defined in \eqref{eq:aux1_def}.

\subsubsection{Nominal: Analysis}

Before providing the main theorem, we state two lemmas that are used in proving the theorem.

\begin{lemma}
(Refer \cite{Cheney1959}) Using the inequality from projections on closed convex sets, for $\Omega  \subset R^n, y\in R^n, \omega\in \Omega$ we have-
\[
(\mathcal{P}_\Omega(y)-w)^T(y-\mathcal{P}_\Omega(y)) \geq 0
\]
\end{lemma}


\begin{lemma}
For displacement-based formation control, with $\Phi_{ij} = \frac{\| p_i - p_j -d_{ij}\|^2}{r^2-\| p_i - p_j\|^2+\mu}$, consider the following function \\
$f = \frac{1}{2}\sum_{i=1}^{N}\sum_{j \in \mathcal{N}_i}\Phi_{ij}$. Then $\dot{f}$ is given by $\sum_{i=1}^{N}\sum_{j \in \mathcal{N}_i}v_i^T\nabla_{p_i}\Phi_{ij}$.
\label{lemma_pot}
\end{lemma}
\begin{proof}
The proof is presented in Appendix.
\end{proof}
The lemma \ref{lemma_pot} establishes the gradient of the combined potential function of the whole multi-agent system. Now we present our main result.


\begin{theorem}
Consider a MAS with $N$ mobile agents and $|\LL|$ stealthy leaders, all with limited sensing region of radius $r$ following the system dynamics given by \eqref{eq:dynamics}. If the initial network $(t=0)$ formed by the agents is connected and each agent starting with zero initial velocity is provided with the proposed control law \eqref{eq:nom_control}-\eqref{eq:aux2_def}, then the following hold
\begin{enumerate}
    \item Network will remain connected for all $t \geq 0$,
    \item Velocity of all agents will converge asymptotically to the same value and will track the reference velocity $v_d$ after reaching the target region,
    \item Asymptotically all the agents converge to the desired formation shape,
    \item The Position of all the stealthy leaders will lie inside the target region as $t \to \infty$. 
\end{enumerate}
\end{theorem}
\begin{proof}
The proof is presented in the Appendix.
\end{proof}
\begin{remark}
    The control scheme \eqref{eq:nom_control}, utilizes two auxiliary variables $\phi_{ij}$ and $\gamma_i$. Here $\phi_{ij}$ is used to estimate the relative velocity difference to perform velocity consensus, on the other hand, $\gamma_i$ drives the velocity of stealthy leaders to track the reference velocity.
\end{remark}

In the absence of tracking velocity, the proposed nominal control gets simplified to
\begin{equation}
\begin{split}
 \bar u_i = &-c_1\sum_{j \in \mathcal{N}_i}\nabla_{p_i}\Phi_{ij} -c_2\sum_{j \in \mathcal{N}_i}\hat v_{ij} \\ &- c_3 \mathbb{1}_{i \in \LL} \mathbb{1}_{\Omega^-} \frac{p_i-\mathcal{P}_\Omega(p_i)}{\|p_i-\mathcal{P}_\Omega(p_i)\|},
\label{eq:nom_control_2}
\end{split}
\end{equation}
where $\mathbb{1}_{i \in \LL}$ is an indicator function whose value is $1$ if agent $i$ is a stealthy leader and $\mathbb{1}_{\Omega}$ indicator functions whose value is $1$ if agent $i$ is inside the target region. The estimation of the relative velocity vector $\hat v_{ij}$ and the auxiliary variable $\phi_{ij}$ is defined in equations \eqref{eq:vel_estimate} and \eqref{eq:aux1_def} respectively. Here  $c_1$, $c_2$, $c_3$,and $\eta$  are strictly positive and $\phi_{ij}(0)$ is initialized with all zeros.

\begin{corollary}
Consider a MAS with $N$ mobile agents and $|\LL|$ stealthy leaders, all with limited sensing region of radius $r$ following the system dynamics given by \eqref{eq:dynamics}. If the initial network $(t=0)$ formed by the agents is connected and each agent is provided with the proposed control law \eqref{eq:nom_control_2}, then the following hold
\begin{enumerate}
    \item Network will remain connected for all $t \geq 0$,
    \item Asymptotically velocity of all agents converges to the same value,
    \item Asymptotically all the agents converge to the desired formation,
    \item Finally, as $t \to \infty$ position of all the stealthy leaders, will lie inside the target region.
\end{enumerate}
\end{corollary}
\begin{proof}
The proof is presented in the Appendix.
\end{proof}

As mentioned earlier, the nominal control laws formulated in this section do not take any safety into account. To guarantee safety, the key idea is to change the nominal control laws in a minimally invasive fashion to ensure that the motion of the agents is safe invariant. 

\subsection{Safe Control}
\label{sec:safe_control}
 
The safety of a multi-agent system can be divided into two components: internal safety and external safety. Internal safety focuses on preventing collisions among the agents, while external safety pertains to safe maneuvering around unsafe regions denoted as $\overline \Omega$. Previous literature often incorporates additional terms in the nominal control law or updates the artificial potential functions to enforce safety. However, safety should be an inherent property of the system, regardless of the objectives being pursued. Inspired by the concepts of barrier certificates and control barrier functions (CBFs) \cite{Wieland2007,Ames2019}, we treat safety as a low-level objective and utilize CBF theory to address it. CBFs provide a more relaxed framework compared to the restrictive conditions of Lyapunov functions, ensuring the system's safe invariance.

Consider a safe set $\mathcal{C}$ in the space $\mathcal{D} \subseteq \mathbb{R}^n$, and a continuously differentiable function $h$ which defines the super level set over $\mathcal{C}$ such that
\begin{equation}
  \mathcal{C} = \{x \in D \subset \mathbb{R}^n : h(x) \geq 0\}.  
  \label{cbf_def}
\end{equation}

Let us consider a generic dynamical system given by
\begin{equation}
\Dot{x} = f(x) + g(x)u.
\label{gen_dyn}
\end{equation}

The treatment of function $h$ as defined above is similar to that of the Lyapunov function, except that here we look for its time derivative to be greater than a certain value, which is proportional to function $h$ with a negative coefficient. Formally, $h$ is a control barrier function (CBF) if there exists an extended class $\kappa_{\infty}$ function $\alpha$ such that for the above control system, the following holds
\begin{equation}
    \sup_{u \in U}\dot{h}(x) = \sup_{u \in U}[L_f h(x) + L_g h(x)u] \geq -\alpha(h(x)) \; \forall \; x \in \mathcal{D}. 
    \label{cbf_1}
\end{equation}
where $L_f h(x)$ and $L_g h(x)$ denote the Lie derivative of $h(x)$ with respect to function $f$ and $g$ respectively. This motivates to design a controller which ensures the above inequality to hold, giving the safety condition as
\begin{equation}
    K_{cbf}(x) = \{u \in U : L_f h(x) + L_g h(x)u +\alpha(h(x)) \geq 0 \}.
\end{equation}

\begin{theorem}
[refer \cite{Ames2019}] Let $\mathcal{C} \subset \mathbb{R}^n$ be a set defined as the super level set of a continuously differentiable function $h : D \subset \mathbb{R}^n$. If $h$ is a control barrier function on $D$ and $\frac{\partial h}{\partial x} (x) \neq 0$ for all $x \in \partial \mathcal{C}$\footnote{The set $\partial \mathcal{C}$ is defined as $\partial \mathcal{C} = \{x \in D \subset \mathbb{R}^n : h(x) = 0\}.$}, then any Lipschitz continuous controller $u(x) \in  K_{cbf} (x)$ for the system (\ref{gen_dyn}) renders the set $\mathcal{C}$ safe. Additionally, the set $\mathcal{C}$ is asymptotically stable in D.
\end{theorem}

One of the limitations of the CBF defined by (\ref{cbf_1}) is that it requires the control input to appear in the first derivative of CBF which leads to complex CBFs for higher-order systems. The same approach can be extended to CBFs with arbitrarily high relative-degree. Such functions are known as exponential control barrier functions (ECBF). The $\rho$th derivative of $h(x)$ can be written as
\begin{equation}
    h^\rho(x) = L_f^\rho h(x) + L_gL_f^{\rho-1}h(x)u.
\end{equation}
We define $\eta_b$ as-
\begin{equation}
    \eta_b(x) := [h(x) \; \dot{h}(x) ...\; h^{\rho-1}(x)]^T.
\end{equation}

A $\rho$ times continuously differentiable function $h$ as defined by (\ref{cbf_def}) is an ECBF if there exists a row vector $K_\alpha \in \mathbb{R}^n$ such that for the above control system (\ref{gen_dyn}), the following holds:
\begin{equation}
    \sup_{u \in U}[L_f^\rho h(x) + L_g L_f^{\rho-1} h(x)u] \geq -K_\alpha \eta_b(x)  \; \forall \; x \in \mathcal{D}. 
\end{equation}

For the second-order CBF, the forward invariance condition boils down to
\begin{equation}
\Ddot{h}+k_1\dot{h}+k_0h \geq 0,
\label{2eq1}
\end{equation}
which is the same as
\begin{equation}
L_f^2 h(\textbf{x}) + L_g L_f h(\textbf{x})u+ k_1 L_f h(\textbf{x}) + k_0 h(\textbf{x}) \geq 0,
\label{cbf_condition}
\end{equation}
where the parameters $k_0$, $k_1$ should be selected such that the two roots of the polynomial $s^2+ k_1s+k_0$ are negative real\footnote{These parameters $k_1$ and $k_0$ determine the extent to which the agents can navigate in proximity to unsafe regions.}.

\begin{figure}[t]
\includegraphics[scale=0.45]{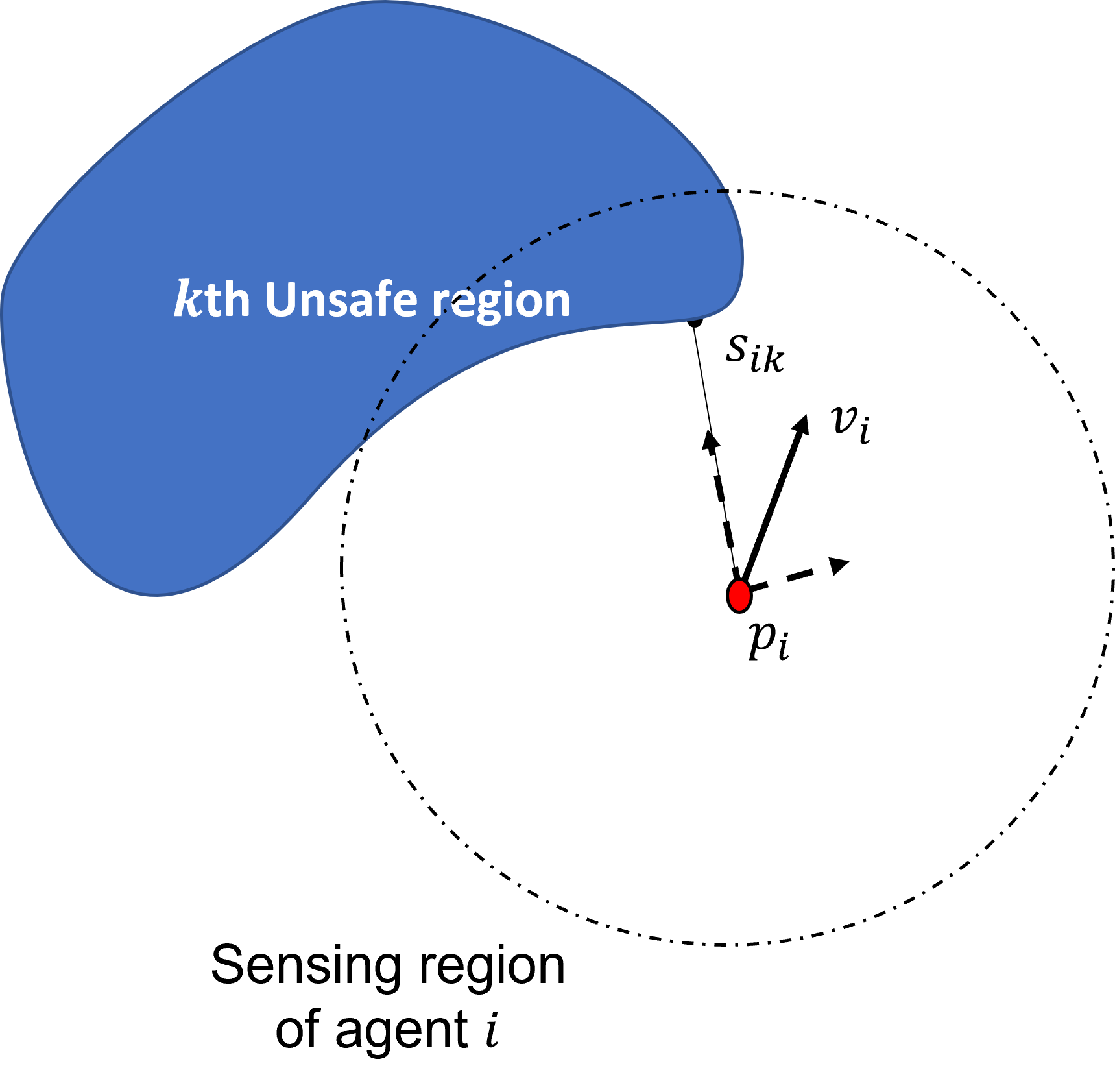}
\centering
\caption{Sensing unsafe region: $s_{ik}$ is closest point on $k$th unsafe region from agent $i$'s position.}
\label{unsafe_fig}
\end{figure}

Let $s_{ik}$ be the closest point from the position of agent $i$ ($p_i$), to the $k$th unsafe region inside the sensing region of agent $i$ (partial or complete) as shown in figure (\ref{unsafe_fig}). Using $\delta_{ex}$ ($\ll r$) as the required safe distance for external safety, we define ECBF for interaction between agent $i$ and $k$th unsafe region as
\begin{equation}
h_{ik}=\| p_i-s_{ik}\|^2-\delta_{ex}^2.
\end{equation}

To ensure safe invariance, we want (\ref{cbf_condition}) to hold true for our given ECBF, which results in the following inequality
\begin{align*}
2v_i^Tv_i + 2(p_i-s_{ik})^Tu_i &+ 2k_1(p_i-s_{ik})^Tv_i \\ & + k_0(\|p_i-s_{ik}\|^2-\delta_{ex}^2) \geq 0.
\end{align*}
This can be simplified and expressed as
\begin{equation}
 A_{ik}+B_{ik}^Tu_i \geq 0,
 \label{eq:cbf_inequality_1}
\end{equation}
where $A_{ik} = 2v_i^Tv_i +  2k_1(p_i-s_{ik})^Tv_i+ k_0(\|p_i-s_{ik}\|^2-\delta_{ex}^2)$ and $B_{ik} = 2(p_i-s_{ik})$.

A similar construction of ECBF is implemented to ensure internal safety, i.e. for collision avoidance within the agents. Consider the interaction between agent $i$ and $j$, then with the required safe distance to be $\delta_{in}$ ($\ll r$), we define the ECBF as  
\begin{equation*}
h_{ij}=\|p_i-p_j\|^2-\delta_{in}^2 \; \forall  \; j \in \mathcal{N}_i; \; i \in \mathcal{V}.
\end{equation*}
As earlier, using equation (\ref{cbf_condition}) we get the inequality as
\begin{equation}
 \bar A_{ij}+ \bar B_{ij}^T(u_i-u_j) \geq 0 \; \forall  \; j \in \mathcal{N}_i; \; j\neq i,
 \label{eq:cbf_inequality_2}
\end{equation}
where $\bar A_{ij} = 2(v_i-v_j)^T(v_i-v_j) +  2k_1(p_i-p_j)^T(v_i-v_j)+ k_0(\|p_i-p_j\|^2-\delta_{in}^2)$ and $\bar B_{ij} = 2(p_i-p_j)$. 

One can define a pairwise safe set $\mathcal{C}_{ij}$ in the space $\mathcal{D} \in \mathbb{R}^n$ as
\begin{equation*}
\mathcal{C}_{ij} = \{(p_i,v_i) \;|\; h_{ik}>0;h_{ij}>0 \} \; \forall i\neq j.
\end{equation*}

We note that direct implementation of the constraints \eqref{eq:cbf_inequality_2} poses two difficulties, it depends on the velocity and control input of both agents which are not shared during the communication. For the velocity dependence, we utilize the velocity estimate $\hat v_{ij}$ defined in \eqref{eq:vel_estimate} by taking into account the error bound of this estimate.

\begin{lemma}
    For the dynamics defines by \eqref{eq:dynamics}, the error bound, $e$, between $(v_i-v_j)$ and the relative velocity estimate $\hat v_{ij}$ defined by \eqref{eq:vel_estimate} is given by
    \begin{align}
    \|e\| &\leq \frac{2u_{max}}{\eta}. 
\end{align}
\label{lemma3}
\end{lemma}

Using Lemma \ref{lemma3}, we observe that relative velocity can be bounded using
\begin{align*}
    \|v_i-v_j\| \geq \|\hat v_{ij}\| - \|e\|. 
\end{align*} 
Whereas the term $(p_i-p_j)^T(v_i-v_j)$ in $\bar A_{ij}$ can be upper bounded using Cauchy-Schwarz inequality as 
\begin{align*}
    (p_i-p_j)^T(v_i-v_j) &= (p_i-p_j)^T(\hat v_{ij} - e),\\
    &\geq (p_i-p_j)^T\hat v_{ij} - \|p_i-p_j\|\|e\|,\\
    &\geq (p_i-p_j)^T\hat v_{ij} - \frac{2\|p_i-p_j\|u_{max}}{\eta}.
\end{align*}
Hence \eqref{eq:cbf_inequality_2} can be lower bounded using the constraint
\begin{equation}
 \tilde A_{ij}+ \bar B_{ij}^T(u_i-u_j) \geq 0 \; \forall  \; j \in \mathcal{N}_i; \; j\neq i,
 \label{eq:cbf_inequality_3}
\end{equation}
where $\tilde A_{ij} = 2(\|\hat v_{ij}\| - \|e\|)^2 +  2k_1(p_i-p_j)^{T} \hat v_{ij} - 2k_1\frac{2\|p_i-p_j\|u_{max}}{\eta} + k_0(\|p_i-p_j\|^2-\delta_{in}^2)$ and $\bar B_{ij} = 2(p_i-p_j)$. Next, in order to address the dependence of \eqref{eq:cbf_inequality_2} on the control input of other agents, we use the decentralized strategy proposed in  \cite{Wang2017} where each agent takes only a fraction of the responsibility  for collision avoidance. Using this, the constraint for agent $i$ can be obtained as
\begin{equation}
\frac{1}{2} \tilde A_{ij}+\bar B_{ij}^Tu_i \geq 0 \; \forall  \; j \in \mathcal{N}_i; \; j\neq i,
\label{eq:cbf_inequality_4}
\end{equation}
where agent $i$ assumes only half the responsibility for avoiding collisions.

Next, we provide the safety guarantee by the following result.

\begin{theorem}
    Consider a MAS with N mobile agents following the system dynamics given by \eqref{eq:dynamics}. If the control law satisfies the constraints \eqref{eq:cbf_inequality_1} and \eqref{eq:cbf_inequality_4} for all agents $i\in \mathcal{V}$, then the MAS is guaranteed to be safe.
\end{theorem}
\begin{proof}
The proof is presented in the Appendix.
\end{proof}

We note that it is not difficult to create complex unsafe sets and deadlocks (either using highly non-convex regions or unsafe regions very close to each other) where the MAS will not be able to achieve the high-level objective, i.e. either reaching the target region or attaining the desired formation. This is because the agents are primarily driven towards the target region using the linear distance between the stealthy leaders and the region $\Omega$, and a big complex unsafe set can directly obstruct the path. It is important to note that the current approach provides a safe maneuvering strategy and does not include path planning. 

\begin{figure}[t]
\includegraphics[width=0.8\linewidth, trim={0.3cm 0.31cm 0.2cm 0.1cm}, clip]{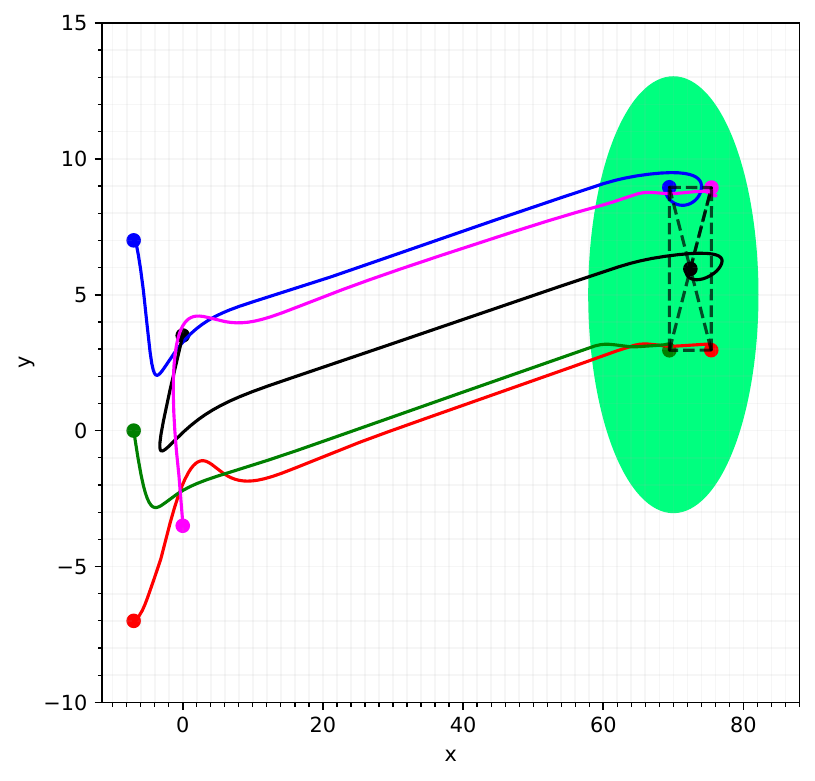}
\centering
\caption{Trajectory of nominal control law for RFC problem without reference velocity tracking. The desired formation is represented by the dashed lines.}
\label{fig:nominal_trajectory_without_track}
\end{figure}

\begin{figure}[t]
\includegraphics[width=0.8\linewidth, trim={0.3cm 0.31cm 0.2cm 0.1cm}, clip]{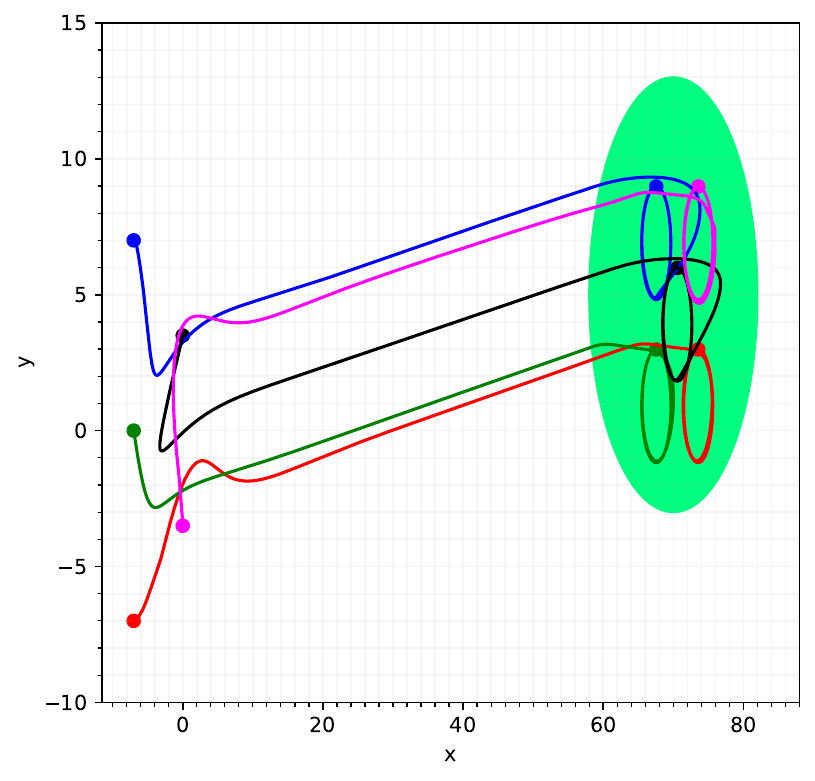}
\centering
\caption{Trajectory of nominal control law for RFC problem with reference velocity tracking.}
\label{fig:nominal_trajectory}
\end{figure}

\begin{figure}[!ht]
\includegraphics[width=\linewidth, trim={0.2cm 0.31cm 0.1cm 0.2cm}, clip]{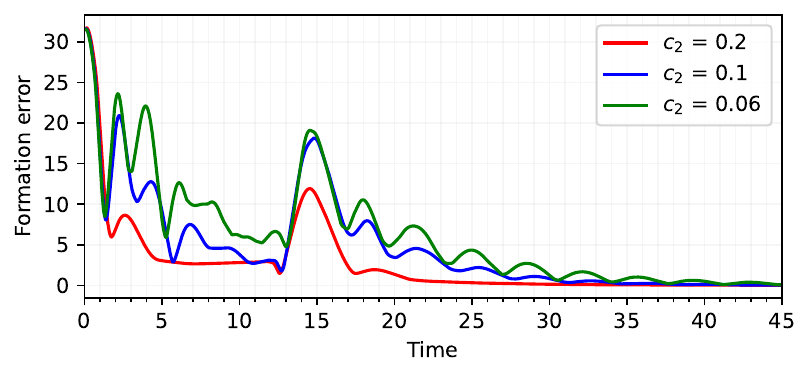}
\centering
\caption{Comparison of formation error for different values of velocity consensus coupling parameter.}
\label{fig:nominal_error}
\end{figure}

\begin{figure*}[!htb]
\includegraphics[width=0.95\linewidth, trim={0.1cm 6.55cm 0.3cm 1.65cm}, clip]{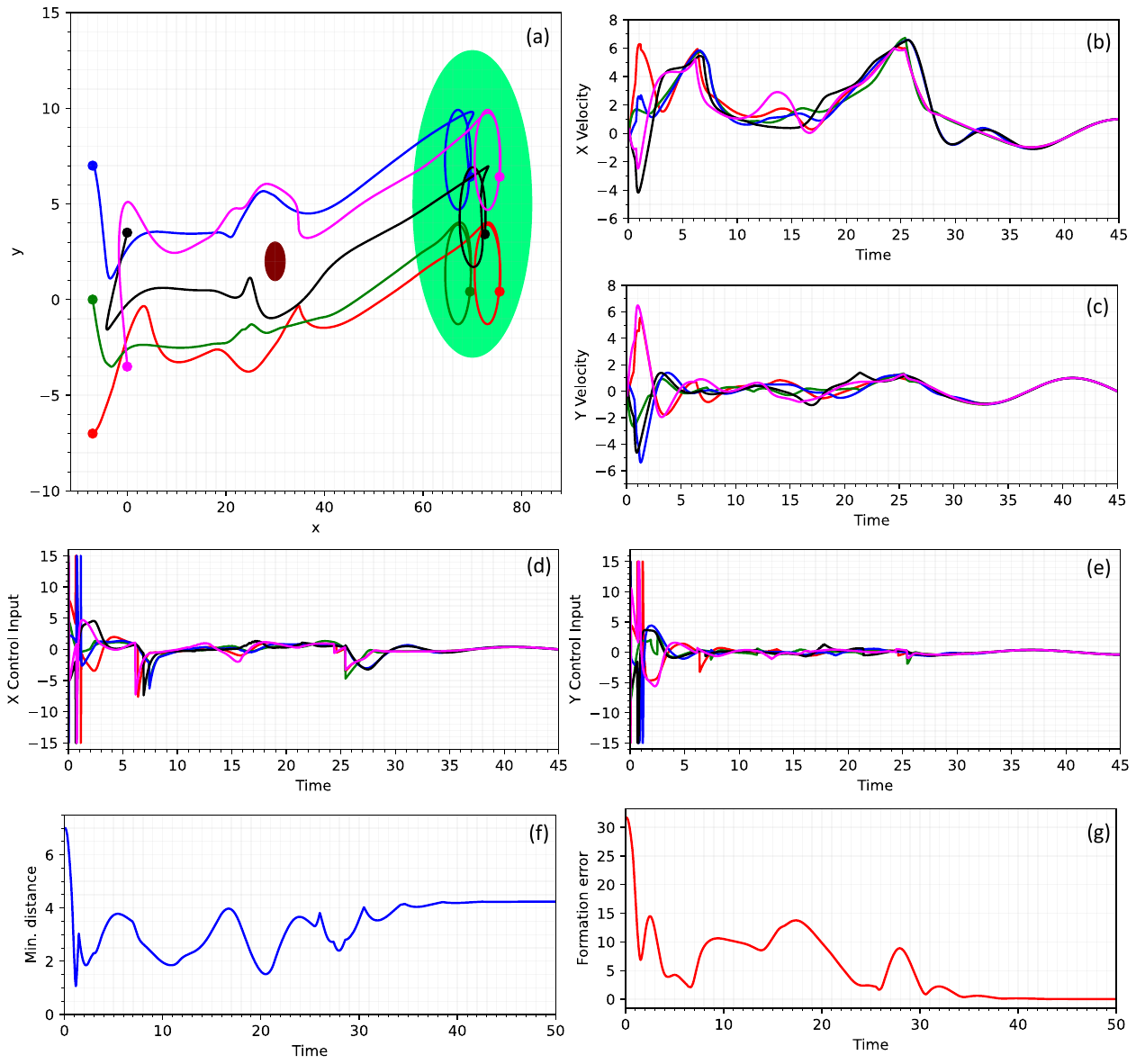}
\centering
\caption{Performance of safe control law for SRFC problem: (a) Trajectory of the agents, (b)-(c) $x$ and $y$ velocity of the agents, (d)-(e) $x$ and $y$ control input of the agents, (f) minimum distance among all the agents, (e) network formation error.}
\label{fig:unsafe_1}
\end{figure*}

\begin{figure*}[!ht]
\includegraphics[width=0.95\linewidth, trim={0.3cm 5.55cm 0.05cm 2.25cm}, clip]{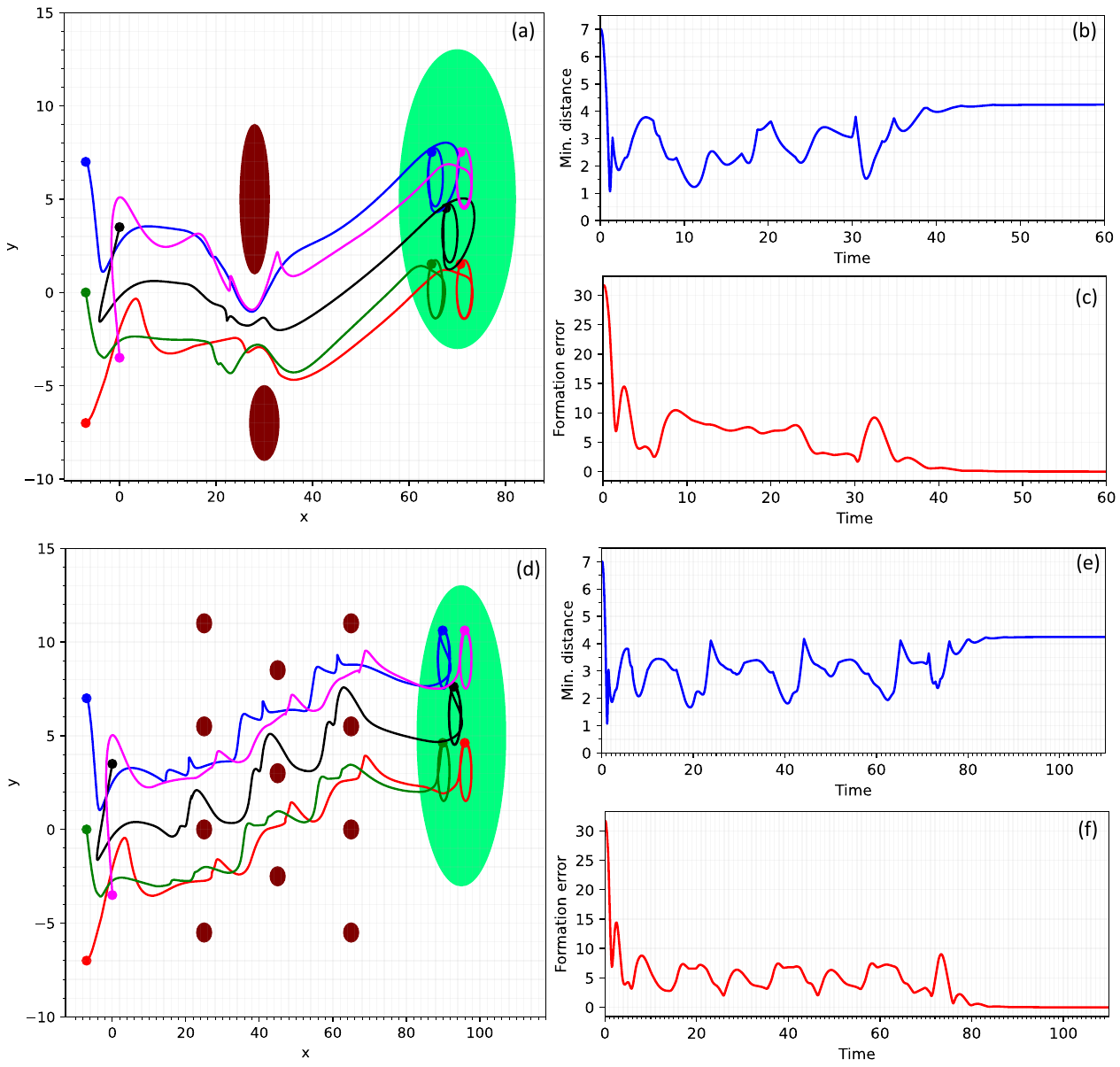}
\centering
\caption{Trajectory and Sensing unsafe region: $s_{ik}$ is closest point on $k$th unsafe region from agent $i$'s position}
\label{fig:unsafe_2}
\end{figure*}

\section{Simulations} 
\label{sec:simulation}
\subsection{Nominal control}
First, we assess the performance and results of the nominal control law. We simulate a group of five mobile agents initially connected at time $t=0$. The target region, denoted as $\Omega$, is depicted in green. The desired formation takes the shape of a rectangle, with the black agent in the center as shown using the dashed line in Fig. \ref{fig:nominal_trajectory_without_track}. For our simulations, we employ a tracking velocity described by $v_d = [v_o \cos(\theta t), v_o \sin(\theta t)]^T$, where $v_o$ and $\theta$ are hyperparameters. All agents start with zero velocity at $t=0$. The parameter values used are $15, 0.2, 1.5, 5, 4,$ and $100$ for $c_1$, $c_2$, $c_3$, $c_4$, $c_5$, and $\eta$, respectively, for all agents in $\mathcal{V}$. Figure \ref{fig:nominal_trajectory_without_track} and \ref{fig:nominal_trajectory} displays the complete trajectory of the agents without and with reference velocity tracking respectively. With the control law defined in equation \eqref{eq:nom_control}, we observe that the multi-agent system (MAS) successfully converges to the desired formation within the specified region. Additionally, it begins to track the reference velocity, which in this case is an elliptical motion. These results align with our theoretical expectations.
    
To highlight the importance of the velocity consensus factor, we compare the formation error for different values of $c_2$. The formation error is defined as
\begin{align*}
    e_f = \sum_{i,j\in \mathcal{V}}  \|p_i-p_j-d_{ij}\|^2,
\end{align*}
which is zero only when the desired formation is achieved. Given the second-order dynamics, every agent exhibits inertia. The potential term provides acceleration to the agent until the desired displacement is not achieved. But by the time the desired displacement is achieved, the agent gains certain momentum which causes it to move beyond the desired equilibrium point. If we consider a two-agent system, then with $c_2 =0$, we observe that they will keep oscillating back and forth without ever converging to a single point. Hence the velocity consensus terms help the agents to drive to the desired formation in a stable fashion. The parameter $c_2$ also attempts to provide a group velocity to the complete formation. As shown in Fig. \ref{fig:nominal_error}, for $c_2 = 0.06$, $0.1$, and $0.2$, we see that with lower values of $c_2$, the agents tend to oscillate. A lower value of $c_2$ also increases the chances of internal collision among the agents.

\subsection{Safe control}
Subsequently, we conduct simulations in the presence of unsafe regions, indicated by the dark red regions. Similar to the previous scenario, all agents establish a connected graph with an initial velocity of zero. The parameter values for $c_1$-$c_5$, and $\eta$ remain consistent with those in Section 4.1. For both internal and external safety, we employ values of $k_1$ and $k_0$ set at 5 and 1, respectively. The safe distance used for external safety is denoted as $\delta_1 = 0.5$, while for internal safety, it is $\delta_2 = 0.1$. In Fig. \ref{fig:unsafe_1}a, the trajectory of the agents in the presence of a single unsafe region is displayed, illustrating how the agents deviate from their desired formation to circumvent the unsafe region. Figures \ref{fig:unsafe_1}b-e present velocity and control input plots, showcasing how relative velocity converges to zero and follows the reference velocity. It's important to note that the control input remains within permissible limits throughout. To ensure that internal collisions are avoided, we plot the minimum distance among all agents over time, as depicted in Fig. \ref{fig:unsafe_1}f. Additionally, the evolution of the formation error is illustrated in Fig. \ref{fig:unsafe_1}e.

We now move on to more complex scenarios involving multiple unsafe regions. In contrast to the trajectory behavior observed when dealing with a single unsafe region, the presence of multiple closely positioned unsafe regions leads to a noticeable adjustment in the formation, as evident in Fig. \ref{fig:unsafe_2}. In the case involving two unsafe regions, as depicted in Fig. \ref{fig:unsafe_2}a, the formation adapts by compressing to create a path through a narrow window, allowing it to avoid the unsafe regions. Figure \ref{fig:unsafe_2}d showcases the trajectory of agents in the presence of multiple closely spaced unsafe regions. To facilitate smoother maneuvering, we relaxed the hyperparameter $c_1$, associated with the potential function, to a value of $1.25$. This adjustment allows the agents to adapt by expanding and contracting their formation, ensuring a safe passage around the unsafe regions. Minimum distance plots for both scenarios are presented in Figure \ref{fig:unsafe_2}b and Figure \ref{fig:unsafe_2}f, confirming the maintenance of internal safety in all cases, whereas formation error plots for both scenarios are presented in Figure \ref{fig:unsafe_2}c and Figure \ref{fig:unsafe_2}g. The closest proximity between agents occurred when navigating around multiple unsafe regions, with a distance of $1.15$ units. It's worth noting that as the number of unsafe regions increases, the trajectory profiles become increasingly complex and require more time to achieve the specified objectives.

\section{Conclusion and Future work} \label{Sec_Con}
This work addresses the problem of safe region formation control (SRFC). The solution is approached by first designing a nominal control to guarantee the region formation control in a limited sensing setup by preserving the connectivity of the network. To deal with safety, instead of modifying the nominal control, safety constraints are imposed by framing a QP problem using ECBFs. This ensures that safety is ensured irrespective of the high-level abstract objective in a minimally invasive fashion. The formulation introduces the stealthy leader as the agents having the information about the target region and the reference tracking velocity signal, but are unidentifiable to other agents. The theoretical and numerical results demonstrate the convergence of the MAS to obtain the desired formation inside the target region, in the presence of unsafe regions. The future work will include: (i) Solving the SRFC problem in the presence of a dynamic unsafe region, (ii) Integrating a high-level trajectory planner for efficient navigation through congested unsafe areas, and (iii) Developing a robust algorithm to counter failures in relative position sensing and errors, ensuring the stability of the SRFC system. 

\appendix
\section{Proof of Lemma 2}
\label{sec:proof_lem1}
\begin{proof}
Let $\lambda_1 = (r^2-\| p_i - p_j\|^2+\mu)$ and $\lambda_2 = (\| p_i - p_j -d_{ij}\|^2)$. As $\| p_i - p_j\|<r$ always hold true, $\lambda_1$ and $\lambda_2$ are always non-negative. Then $\dot{f}$ is given by- 
\begin{align*}
\dot{f} &= \frac{1}{2}\sum_{i=1}^{N}\sum_{j \in \mathcal{N}_i}\dot{\Phi}_{ij}, \\
&= \frac{1}{2}\sum_{i=1}^{N}\sum_{j \in \mathcal{N}_i} \left(v_i-v_j\right)^T\left(\frac{2\lambda_1(p_i - p_j -d_{ij})
+ 2\lambda_2(p_i - p_j)}{(\lambda_1)^2}\right), \\
&= \frac{1}{2}\sum_{i=1}^{N}\sum_{j \in \mathcal{N}_i} v_i^T\left(\frac{2\lambda_1(p_i - p_j -d_{ij})
+ 2\lambda_2(p_i - p_j)}{(\lambda_1)^2}\right) \\ & + \frac{1}{2}\sum_{i=1}^{N}\sum_{j \in \mathcal{N}_i} v_j\left(\frac{2\lambda_1(p_j - p_i -d_{ji})
+ 2\lambda_2(p_j - p_i)}{(\lambda_1)^2}\right).
\end{align*}
As we have $d_{ij} = -d_{ji}$, we get 
\begin{align*}
&= \sum_{i=1}^{N}\sum_{j \in \mathcal{N}_i} v_i^T\left(\frac{2\lambda_1(p_i - p_j -d_{ij})
+ 2\lambda_2(p_i - p_j)}{(\lambda_1)^2}\right), \\
&= \sum_{i=1}^{N}\sum_{j \in \mathcal{N}_i}v_i^T\nabla_{p_i}\Phi_{ij}.
\end{align*} 
This completes the proof of the lemma.
\end{proof}


\section{Proof of Theorem 1}
\label{sec:proof_thm1}

\begin{proof}
We define the velocity tracking signal as
\begin{align}
    \Tilde{v}_i = v_i- \mathbb{1}_{\Omega}v_d 
    \label{eq:v_tilde}
\end{align}
Consider the following Lyapunov function for some $\alpha>0$:
\begin{equation}
\begin{aligned}
V = & \underbrace{\frac{1}{2}\sum_{i=1}^{N}\Tilde{v}_i^T \Tilde{v}_i}_{V_1} +  \underbrace{\frac{c_1}{2}\sum_{i=1}^{N}\sum_{j \in \mathcal{N}_i}\Phi_{ij}}_{V_2}  +  \underbrace{c_3\sum_{i=1}^{N}\mathbb{1}_{i \in L}\mathbb{1}_{\Omega^-}\|p_i-\mathcal{P}_\Omega(p_i)\|}_{V_3} \\ 
& + \underbrace{\frac{c_4}{c_5}\sum_{i=1}^{N} \mathbb{1}_{i \in L}\mathbb{1}_{\Omega} \boldsymbol{1}_n^T \log(\cos(c_5(p_i-\gamma_i)))}_{V_4}+  \underbrace{\frac{\zeta}{2}\sum_{i=1}^{N}\sum_{j \in \mathcal{N}_i}\dot \phi_{ij}^T \dot \phi_{ij}}_{V_5}.
\end{aligned}
\end{equation}

The time derivative $V$ for the system dynamics given by \eqref{eq:dynamics} can be decomposed by

\begin{equation}
    \dot{V} = \dot{V}_1 + \dot{V}_2+\dot{V}_3+\dot{V}_4+\dot{V}_5
\label{eq:thm1_V_dot}
\end{equation}

The derivative of $V_1$ can be obtained using \eqref{eq:v_tilde} and control input \eqref{eq:nom_control} as
\begin{align*}
\dot{V}_1 &= \sum_{i=1}^{N}\Tilde{v}_i^T (u_i-\mathbb{1}_{\Omega}\dot{v}_d-\dot{\phi_i})\\
&= \sum_{i=1}^{N}\Tilde{v}_i^T \Big(-c_1\sum_{j \in \mathcal{N}_i}\nabla_{p_i}\Phi_{ij} -c_2\sum_{j \in \mathcal{N}_i}\hat v_{ij} \\
&- c_3 \mathbb{1}_{i \in L}\mathbb{1}_{\Omega^-} \frac{p_i-\mathcal{P}_\Omega(p_i)}{\|p_i-\mathcal{P}_\Omega(p_i)\|} -c_4\mathbb{1}_{i \in L}\mathbb{1}_{\Omega} \tanh(c_5(p_i-\gamma_i))\Big).
\end{align*}

Using the lemma 1, the derivative of $V_2$ is 
\begin{align*}
\dot{V}_2 &= c_1\sum_{i=1}^{N}\sum_{j \in \mathcal{N}_i}v_i^T \nabla_{p_i}\Phi_{ij}\\
&= c_1\sum_{i=1}^{N}\sum_{j \in \mathcal{N}_i}\Tilde{v}_i^T \nabla_{p_i}\Phi_{ij} \hspace{10pt} (\text{As $v_i-v_j = \Tilde{v}_i -\Tilde{v}_j$}).
\end{align*}

By leveraging the property that the derivative of a projection aligns with the tangent of the convex region, we derive the expression for $\dot{V}_3$ as follows
\begin{align*}
\dot{V}_3 &= c_3\sum_{i=1}^{N} \mathbb{1}_{i \in L}\mathbb{1}_{\Omega^-} v_i^T\frac{p_i-\mathcal{P}_\Omega(p_i)}{\|p_i-\mathcal{P}_\Omega(p_i)\|}\\
&= c_3\sum_{i=1}^{N} \mathbb{1}_{i \in L}\mathbb{1}_{\Omega^-} (\Tilde{v}_i+ \mathbb{1}_{\Omega}v_d)^T\frac{p_i-\mathcal{P}_\Omega(p_i)}{\|p_i-\mathcal{P}_\Omega(p_i)\|}\\
&= c_3\sum_{i=1}^{N} \mathbb{1}_{i \in L}\mathbb{1}_{\Omega^-} \Tilde{v}_i^T\frac{p_i-\mathcal{P}_\Omega(p_i)}{\|p_i-\mathcal{P}_\Omega(p_i)\|}, \hspace{25pt}
\end{align*}

Derivative of $V_4$ can be simplified using \eqref{eq:v_tilde}, \eqref{eq:aux2_def} as
\begin{align*}
\dot{V}_4 &= c_4\sum_{i=1}^{N} \mathbb{1}_{i \in L} \mathbb{1}_{\Omega}(v_i-\dot{\gamma}_i)^T \tanh(c_5(p_i-\gamma_i))\\
&= c_4\sum_{i=1}^{N} \mathbb{1}_{i \in L} \mathbb{1}_{\Omega} (\Tilde{v}_i+\mathbb{1}_\Omega v_d-\dot{\gamma}_i)^T \tanh(c_5(p_i-\gamma_i))\\
&= c_4\sum_{i=1}^{N} \mathbb{1}_{i \in L} \mathbb{1}_{\Omega} (\Tilde{v}_i-c_5(p_i-\gamma_i))^T \tanh(c_5(p_i-\gamma_i)).
\end{align*}

Lastly, derivative of $V_5$ can be simplified using \eqref{eq:aux1_def} as
\begin{align*}
\dot{V}_5 &= \zeta\sum_{i=1}^{N}\sum_{j\in \mathcal{N}_i}\dot \phi_{ij}^T \ddot{\phi}_{ij} \\
&= \zeta\sum_{i=1}^{N}\sum_{j\in \mathcal{N}_i}\dot \phi_{ij}^T (-\eta(\dot \phi_{ij} - (v_i-v_j)))
\\
&= -\zeta\eta\sum_{i=1}^{N}\sum_{j\in \mathcal{N}_i}\dot \phi_{ij}^T\dot \phi_{ij} +\zeta\eta\sum_{i=1}^{N}\sum_{j\in \mathcal{N}_i} \hat v_{ij}^T(v_i-v_j)\\
&= -\zeta\eta\sum_{i=1}^{N}\sum_{j\in \mathcal{N}_i}\dot \phi_{ij}^T\dot \phi_{ij} +2\zeta\eta\sum_{i=1}^{N}\tilde v_i^T\sum_{j\in \mathcal{N}_i} \hat v_{ij}.
\end{align*}

Using equation \eqref{eq:thm1_V_dot} the derivative of the Lyapunov function is
\begin{align*}
\dot{V} &= -(c_2-2\zeta\eta)\sum_{i=1}^{N} \Tilde{v}_i^T\sum_{j \in \mathcal{N}_i}\hat v_{ij} -\zeta\eta\sum_{i=1}^{N}\sum_{j\in \mathcal{N}_i}\dot \phi_{ij}^T\dot \phi_{ij}   \\& -c_4 \sum_{i=1}^{N} \mathbb{1}_{i \in L}\mathbb{1}_{\Omega} (c_5(p_i-\gamma_i))^T \tanh(c_5(p_i-\gamma_i)) 
\end{align*}
It is worth noting that the first term will become zero by selecting appropriate values of $\zeta$ (specifically, $\zeta = c_2/2n$) in the construction of a Lyapunov function as both $c_2$ and $\eta$ are positive. This results in
\begin{align}
\dot{V} &= -\zeta\eta\sum_{i=1}^{N}\sum_{j\in \mathcal{N}_i}\dot \phi_{ij}^T\dot \phi_{ij}  \notag \\& -c_4 \sum_{i=1}^{N} \mathbb{1}_{i \in L}\mathbb{1}_{\Omega} (c_5(p_i-\gamma_i))^T \tanh(c_5(p_i-\gamma_i)). \label{thm2_v_dot}
\end{align}
which is negative semi-definite. Hence we note that $v_i$, $\phi_{ij}$, $\Phi$,  and $\mathbb{1}_{i \in L} (p_i-\mathcal{P}_\Omega(p_i))$ for $i \in \mathcal{V}$  are globally bounded. Since the network is assumed to be initially connected, and $\Phi$ is bounded, no edge of the dynamic graph is lost at any time $t \geq 0$ preserving the connectivity of the formation. 

If $\dot V \equiv 0$, this results in $\dot \phi_{ij} = 0$ and $\mathbb{1}_\Omega(p_i-\gamma_i) = 0 \; \forall \;i\in \mathcal{L}$. The former also implies $\ddot \phi_{ij} = 0$, which results in $v_i = v_j$ by differentiating \eqref{eq:aux1_def}. By differentiating $\mathbb{1}_\Omega(p_i-\gamma_i) = 0$ and using \eqref{eq:aux2_def} we get $v_i = \mathbb{1}_\Omega v_d \; \forall \;i\in \mathcal{L}$, and as $ v_i = v_j \; \forall \;i,j\in \mathcal{V}$ indeed $v_i = \mathbb{1}_\Omega v_d \; \forall \;i\in \mathcal{V}$. Hence by LaSalle’s Invariance
principle, we conclude that that $\lim_{t\to \infty}(v_i = v_j = \mathbb{1}_\Omega v_d) \; \forall \; i,j \in \mathcal{V}$.

By taking derivative of $v_i = \mathbb{1}_\Omega v_d \; \forall \;i\in \mathcal{V}$, results in $u_i = \mathbb{1}_\Omega \dot v_d \; \forall \;i\in \mathcal{V}$ and using the nominal control input equation \eqref{eq:nom_control}, we get
\begin{equation}
c_1\sum_{j \in \mathcal{N}_i}\nabla_{p_i}\Phi_{ij} + c_3 \mathbb{1}_{i \in \LL} \mathbb{1}_{\Omega^-} \frac{p_i-\mathcal{P}_\Omega(p_i)}{\|p_i-\mathcal{P}_\Omega(p_i)\|} = 0,
\label{eq:thm1_cond}
\end{equation}
Using the fact that the $v_i$ is bounded and the fact that $\dot{\mathcal{P}}_\Omega(p_i))$ is always bounded given that the target region is a finite space, we observe that $(v_i - \dot{\mathcal{P}}_\Omega(p_i))$ is bounded $\forall \; i \in \mathcal{V}$. From Barb$\Breve{\text{a}}$rat Lemma we conclude that $\lim_{t\to \infty} \mathbb{1}_{i \in L}(p_i - \mathcal{P}_\Omega(p_i)) = 0$ $\forall \; i \in \mathcal{V}$. Hence we formally guarantee that all the stealthy leaders will lie in the target region $\Omega$ as $t \to \infty$. 

Lastly using \eqref{eq:thm1_cond} we obtain that $\lim_{t\to \infty}\sum_{j \in \mathcal{N}_i}\nabla_{p_i}\Phi_{ij} = 0$ which holds true only when $\lim_{t\to \infty} \Phi_{ij} = 0$, i.e. when $\lim_{t\to \infty} (p_i - p_j) = d_{ij}$. Hence asymptotically all the agents converge to the desired formation completing the proof. With $d_{ij} = 0$, we get $p_i = p_j \; \forall \; i,j \in \mathcal{V}$ as $t \to \infty$ which makes it a rendezvous problem.
\end{proof}

\section{Proof of Corollary 1}
\label{sec:proof_col1}
\begin{proof}
Consider the following Lyapunov function
\begin{equation}
\begin{split}
V =& \underbrace{\frac{1}{2}\sum_{i=1}^{N}v_i^T v_i}_{V_1} +  \underbrace{\frac{c_1}{2}\sum_{i=1}^{N}\sum_{j \in \mathcal{N}_i}\Phi_{ij}}_{V_2} + \\ &c_3\underbrace{\sum_{i=1}^{N}\mathbb{1}_{i \in L}\mathbb{1}_{\Omega^-}\norm{p_i - \mathcal{P}_\Omega(p_i)}}_{V_3} +  \underbrace{\frac{\zeta}{2}\sum_{i=1}^{N}\phi_{ij}^T \phi_{ij}}_{V_4}
\end{split}
\end{equation}

By employing analogous reasoning to that presented in Theorem 1, the corollary can be proven. 
\end{proof}


\section{Proof of Lemma 3}
\label{sec:proof_lem3}
\begin{proof}
    Let $e = \hat v_{ij}-(v_i-v_j)$, then
\begin{align*}
    \dot e &= \dot{\hat{v}}_{ij}- (\dot v_i- \dot v_j), \\
    &= -\eta ( \dot \phi_{ij} - (v_i-v_j))- (u_i- u_j), \\
    &= -\eta ( \hat v_{ij} - (v_i-v_j))- (u_i- u_j), \\
    &= -\eta e- (u_i- u_j).
\end{align*}
Hence we get
\begin{align*}
    e &= \exp(-\eta t) e(0)+\int_0^t \exp(-\eta (t-\tau))(u_i- u_j)d \tau  \\
    \|e\| &\leq \exp(-\eta t)\|e(0)\|+\int_0^t \exp(-\eta (t-\tau))\|u_i- u_j\|d \tau  \\
    &\leq \exp(-\eta t)\|e(0)\|+ \frac{2u_{max}}{\eta} 
\end{align*}
Using the assumption that the initial velocity of all the agents is $0$ and setting the initial value of $\phi_{ij}(0) = (p_i(0)-p_i(0))$ we get
\begin{align*}
    \|e\| &\leq \frac{2u_{max}}{\eta}. 
\end{align*}
This completes the proof.
\end{proof}


\section{Proof of Theorem 3}
\label{sec:proof_lem3}
\begin{proof}
If the control law of all agents satisfies the constraints \eqref{eq:cbf_inequality_1} and \eqref{eq:cbf_inequality_4} for all agents $i \in \mathcal{V}$, then $\mathcal{C}{ij}$ is forward invariant for all $j \in \mathcal{N}i$ and $i \in \mathcal{V}$, and for all $k=\{\bar k\in \mathcal{K}|\|p_i-s_{i\bar k}\|<r\}$, which is ensured by the design of the barrier function. Additionally, if $j \notin \mathcal{N}i$ or $\|p_i-s_{ik}\|>r$, the safe distance is much smaller than the sensing region, resulting in $h{ij}>0$ and $h_{ik}>0$, respectively. Hence, $\mathcal{C}_{ij}$ remains forward invariant under these conditions.
\end{proof}

\bibliographystyle{elsarticle-num}
\bibliography{reference}

\end{document}